\documentclass[12pt]{iopart}

\begin{document}

\title[
Near Horizon Extreme Magnetized Kerr Geometry]{Near Horizon Extreme Magnetized Kerr Geometry}

\author{Muhammad F A R Sakti$ ^1 $, Philin Y D Sagita$ ^1 $, A Suroso$ ^{1,2} $ and Freddy P Zen$ ^{1,2} $}

\address{$ ^1 $Theoretical Physics Lab., THEPI Division, and \\ $ ^2 $Indonesia Center for Theoretical and Mathematical Physics (ICTMP), Institut Teknologi
Bandung, Jl. Ganesha 10 Bandung 40132, Indonesia}
\ead{m.fitrah@students.itb.ac.id}
\vspace{10pt}
\begin{indented}
\item[]November 2016
\end{indented}

\begin{abstract}
The conjectured magnetized Kerr/CFT correspondence states that the quantum theory of gravity in the near horizon of extreme Kerr black holes immersed by the magnetic field, Near Horizon Extreme Magnetized Kerr black holes, is holographic dual to a two-dimensional chiral conformal field theory. To obtain Near Horizon Extreme Magnetized Kerr geometry, the extreme limit of the magnetized Kerr metric is taken so, $ a=M $ and then continued by transforming the coordinates to have a warped and twisted product of $ \textrm{AdS}_2 \times \textrm{S}^2 $, and also with the Near Horizon Extreme Kerr metric one. Consequently, we can obtain also the new Ernst potentials for those geometries. Finally, the transformed central charge from the extremal non-magnetized one to the magnetized one in the Ernst-Papapetrou formalism is obtained.
\end{abstract}

%
\hspace{5.5pc}

\section{Introduction}
Black hole solutions are related to the Ernst potentials \cite{ernst,ernst1}. Ernst potentials are the solutions of the mathematical equations called by Ernst equations which are equivalent to the Einstein field equation in general relativity. In Ernst formalism, we use general rotating black hole metric \cite{reina} called by Papapetrou metric which corresponds to the killing vector $ \hat{\xi}_t $ and $ \hat{\xi}_{\phi} $. This metric can be applied to the more general black holes, Kerr-Newman black holes. In our case, the intrinsic charge of the Kerr-Newman spacetime is taken to be zero in order to obtain the Kerr spacetime.

As we have told above, our main topic is about the Kerr spacetime but we are going to specify to the extremal case because this geometry is related to the recently discussed theory which is holographic dualities. Holographic dualities tell the relation between quantum gravity and quantum field theory. One of the applications of this is so-called Kerr/CFT correspondence. The conjectured Kerr/CFT correspondence states that the quantum theory of gravity in the near horizon of extreme black holes is holographic dual to the conformal field theory. It also has been quite successful as the tools to find the microscopic origin of the Bekenstein-Hawking entropy. 
The entropy of the extremal black holes can be calculated, but first, we need to find the central charge. This charge appears from the Cardy formula. To obtain the corresponding central charge, we can adopt the method from Brown and Henneaux \cite{brown} which is associated to $ \textrm{AdS}_3 $ spacetime.

The tools of Kerr/CFT correspondence are also applied to the Kerr black holes which are immersed by the external magnetic field, that also have a warped and twisted product of $ \textrm{AdS}_2 \times \textrm{S}^2 $ such as the extremal Kerr one. The magnetized geometry firstly was proposed by Melvin that solves the Maxwell-Einstein system of equations. In addition, such physical magnetic fields are believed to be in the active galactic nuclei that have an important role in explaining the physical aspects of the supermassive black holes in the center of the galaxies \cite{elvis}. The magnetized geometry also can be obtained from another fashion such as that is proposed by Harrison \cite{harrison} called by Harrison transformation to the Ernst potentials. 

The consequence of applying Harrison transformation to the metric, simultaneously to the Ernst potentials, is the central charge of the Virasoro algebra which generates the asymptotic symmetries of the near horizon geometry are also found to depend on the magnetic fields parameter. Astorino also has applied the Harrison transformation to the Kerr-Newman spacetime and found the extremal case to obtain the central charge \cite{astorino} and by taking the angular momentum per mass to be zero, we can see it reduces to the case of magnetized Reisser-Nordstr{\"o}m \cite{astorino2}, different with \cite{garousi} that extends Reisser-Nordstr{\"o}m to 5D to have off-diagonal term. The Near Horizon Extreme Magnetized Kerr (NHEMK) geometry of the magnetized Kerr black holes has been obtained in \cite{siahaan, astorino} by using Harrison transformation. Finally, the central charge can also be obtained by using stretched horizon formalism \cite{chen} which remains the same with the fashion showed in \cite{hartman}. In addition to the magnetized Kerr/CFT correspondence, the temperature diverges at $ |B| \rightarrow 1/M $ and can even can be negative for the value  $ |B| > M $. Furthermore, in the case of magnetized Kerr-Newman, the central charge is more complicated because of the dependence of the intrinsic charge.

In this following, the Ernst potentials of the extremal case of Kerr and magnetized Kerr are shown by using coordinates transformation such in obtaining the extremal metrics. We can see that the potentials just depend on an angular coordinate in the extremal case. The vector potentials of NHEMK geometry are also obtained as the solution of the Einstein-Maxwell system of equations which correspond to the Cartan components. Note that we have obtained the central charge of extremal geometry, which corresponds to the entropy of the black holes. Then we show the transformation of the central charge of NHEK geometry to NHEMK geometry in the form of Ernst and Papapetrou potentials.

We organize the remaining parts of the paper as follows. In section 2, we review the part of Kerr metric and the magnetized one and its Ernst potentials as the solutions of Ernst equations. In section 3, we investigate the NHEK geometry and its Ernst potentials. The new Ernst potentials of the NHEMK geometry are told in section 4 along with Cartan components. The transformation of the central charge from NHEK geometry to NHEMK one is also shown. Finally, we summarize the whole paper in the last section.

\section{Review on Kerr and magnetized Kerr black hole}

The classical stationary axially symmetric body is described by the Lagrangian density that is independent of time $ t $ and azimuth coordinate $ \phi $. So, it will invariant under these transformations
\begin{eqnarray}
t \rightarrow -t \nonumber\\
\phi \rightarrow -\phi \nonumber .\
\end{eqnarray}
This can be expressed in the form so-called Papapetrou metric \cite{reina} defined by
\begin{equation}
ds^2=f^{-1}\left(-2P^{-2}d\zeta.d\zeta^*+\rho^2d\hat{t}^2\right) -f\left(d\hat{\phi}-\omega d\hat{t}\right)^2 , 
\label{papapetrou}
\end{equation}
and for the case of the metric consists of the radial and angular coordinates, we can choose
\begin{eqnarray}
d\zeta=\frac{1}{\sqrt{2}}\left(\frac{d\hat{r}}{\sqrt{\Delta}}+id\theta\right). \label{papaptambahan}\
\end{eqnarray}
Hence, equation (\ref{papapetrou}) becomes
\begin{equation}
ds^2=f^{-1}\left[-P^{-2}\left(\frac{d\hat{r}^2}{\Delta}+d\theta^2 \right)+\rho^2d\hat{t}^2\right] -f\left(d\hat{\phi}-\omega d\hat{t}\right)^2 , 
\label{papapetrou1}
\end{equation}
and we have chosen that potentials $ f, P, \rho, \omega, \Delta $ in the Papapetrou metric to depend on $ r, \theta $ coordinates. Now, to obtain the Kerr metric \cite{kerr}, we define the potentials to be
\begin{eqnarray}
\rho=\sqrt{\Delta}\mathrm{sin}\theta,~ P=(\sqrt{A}\mathrm{sin}\theta)^{-1}, ~ f=-\frac{A \mathrm{sin}^2\theta}{\Sigma}, ~ \omega=\frac{2Ma\hat{r}}{A}. \label{papapetrouparam}\
\end{eqnarray}
where $ \omega $ is the angular velocity of the black hole. Exactly, we obtain Kerr metric in Boyer-Lindquist coordinates such as
\begin{eqnarray}
ds^2&=&\Sigma\left(-\frac{\Delta}{A}d\hat{t}^2 +\frac{d\hat{r}^2}{\Delta}+d\theta ^2 \right)+\frac{A \mathrm{sin}^2\theta}{\Sigma}\left(d\hat{\phi}-\frac{2M a \hat{r}}{A}d\hat{t}^2\right)^2 ,\ \label{kerrmetric}
\end{eqnarray}
where
\begin{eqnarray}
&&\Delta \equiv \hat{r}^2-2M \hat{r}+a^2, \nonumber \\
&& \Sigma \equiv \hat{r}^2 + a^2 \mathrm{cos}^2\theta, \nonumber \\
&&A \equiv (\hat{r}^2 + a^2)^2-\Delta a^2 \mathrm{sin}^2\theta . \nonumber \
\end{eqnarray}
The constant $ a $ is equal to the angular momentum $ J $ per mass $ M $. 
This equation (\ref{kerrmetric}) is the solution of Ernst equations, defined by
\begin{eqnarray}
(\mathrm{Re} ~ \varepsilon + |\Phi|^2) \nabla ^2 \varepsilon &=& ( \nabla \varepsilon +2 \Phi ^* \nabla \Phi).  \nabla \varepsilon , \label{persamaanernst2} \\
(\mathrm{Re} ~ \varepsilon + |\Phi|^2) \nabla ^2 \Phi &=& ( \nabla \varepsilon +2 \Phi ^* \nabla \Phi).  \nabla \Phi , \label{persamaanernst3} \
\end{eqnarray}
where $\varepsilon = f-|\Phi|^2+i\varphi $. Here, $ \varepsilon $ and $ \Phi $ are called by Ernst potentials. For Kerr black hole, the potentials are
\begin{eqnarray}
\Phi &=& 0, \\
\varepsilon &=& -(\hat{r}^2+a^2)\mathrm{sin}^2\theta +2iMa \mathrm{cos}\theta(3-\mathrm{cos}^2\theta)-\frac{2Ma^2 \mathrm{sin}^4\theta}{\hat{r}+ia \mathrm{cos}\theta}. \label{ernstpotkerr}\
\end{eqnarray}
The event horizon of Kerr black hole is
\begin{eqnarray}
r_\pm = M \pm \sqrt{M^2 -a^2}. \label{kerreventhor}
\end{eqnarray}
In the case of Kerr black hole, extreme limit happens when the constant $ a $ approaches $ M $. Because of this limit, we see that the event horizon become $r_+=M $. In the next section, we will study the geometry of the near horizon $ r=r_+ $ of this extreme limit.

Now, we shift to the magnetized Kerr metric where it is Kerr black hole that is immersed by external magnetic field. This external magnetic field generates non-asymptotically flat black hole solution. Here, we review the usual fashion to obtain this metric by using Harrison transformation \cite{harrison} on the Ernst potentials which are defined by
\begin{eqnarray}
\varepsilon ' = \Lambda^{-1}\varepsilon , ~ \mathrm{dan}~ \Phi ' = \Lambda^{-1}\left(\Phi - \frac{1}{2}B \varepsilon \right) , \ \label{transformasiharrison}
\end{eqnarray}
where
\begin{equation}
\Lambda = 1+ B \Phi - \frac{1}{4} B^2 \varepsilon , \label{LambdaHarrison}\
\end{equation}
$ B $ is the magnetic field parameter. Some Papapetrou potentials also change, those are
\begin{eqnarray}
f' &=& |\Lambda | ^{-2} f, \\
\nabla \omega ' &=& |\Lambda|^2 \nabla \omega - \rho f^{-1}\left(\Lambda^* \nabla \Lambda-\Lambda \nabla \Lambda^* \right),\ \label{transformasiharrisonp}
\end{eqnarray}
but $ \rho$ and $P $ do not change. The Harrison transformation is done on the Kerr metric (\ref{kerrmetric}), hence we obtain magnetized Kerr metric
\begin{equation}
ds^2 = \Sigma |\Lambda |^2 \left(-\frac{\Delta}{A}d\hat{t}^2+\frac{d\hat{r}^2}{\Delta}+d\theta^2 \right)+\frac{A \mathrm{sin}^2\theta}{\Sigma |\Lambda |^2}\left(d\hat{\phi}-\omega' d\hat{t}\right)^2 , \label{melvinkerr}
\end{equation}
where it is the electrovacuum solution of the Einstein field equation. The potential $ \omega ' $ is obtained from (\ref{transformasiharrisonp}) and can be separated to this relations \cite{aliev}
\begin{eqnarray}
\frac{\partial \omega '}{\partial r}&=&|\Lambda |^2 \frac{\partial \omega}{\partial r}-\frac{2\Sigma}{A\mathrm{sin}\theta}\left(\mathrm{Im}~\Lambda\frac{\partial  \mathrm{Re}~\Lambda}{\partial \theta}-\mathrm{Re}~\Lambda\frac{\partial  \mathrm{Im}~\Lambda}{\partial \theta} \right) ,  \\
\frac{\partial \omega '}{\partial \theta}&=&|\Lambda |^2 \frac{\partial \omega}{\partial  \theta}+\frac{2\Sigma \Delta}{A\mathrm{sin}\theta}\left(\mathrm{Im}~\Lambda\frac{\partial  \mathrm{Re}~\Lambda}{\partial r}-\mathrm{Re}~\Lambda\frac{\partial  \mathrm{Im}~\Lambda}{\partial r} \right). \label{integralomega}\ 
\end{eqnarray}
So we will find
\begin{eqnarray}
\omega ' &=& \frac{16 M\hat{r}a + \omega_b B^4}{8A},\\
\omega_b &=& 4a^3M^3\hat{r}(3 + \mathrm{cos}^4\theta) +2aM^2[\hat{r}^4 \{\ (\mathrm{cos}^2\theta - 3)^2 - 6\}\ +2a^2\hat{r}^2(3 \nonumber \\
& &- 3 \mathrm{cos}^2\theta - 2\mathrm{cos}^4\theta) - a^4 (1 + \mathrm{cos}^4\theta)] + aM\hat{r} (\hat{r}^2 + a^2) \{\ \hat{r}^2 (3 \nonumber \\
& &+ 6 \mathrm{cos}^2\theta - \mathrm{cos}^4\theta) - a^2 (1 - 6 \mathrm{cos}^2\theta- 3 \mathrm{cos}^4\theta)\}\ .\
\end{eqnarray}
However, Hiscock \cite{hiscock} have found that magnetized Kerr metric (\ref{melvinkerr}) suffers conical singularity in the axial coordinate $ \hat{\phi} $ hence one rotation of $ \hat{\phi} $ is not $ 2\pi $ anymore but $ 2\pi |\Lambda_0 |^2$  where $ \Lambda_0 $ is equal to $ \Lambda|_{\theta=0} $. In order to solve this problem, we need to scale the axial coordinate, i. e. $ \hat{\phi '} \rightarrow |\Lambda_0 |^2\hat{\phi'} $. Finally, magnetized Kerr metric (\ref{melvinkerr}) will be
\begin{eqnarray}
ds^2&=&\Sigma |\Lambda |^2 \left(-\frac{\Delta}{A}d\hat{t}^2+\frac{d\hat{r}^2}{\Delta}+d\theta^2 \right)+\frac{A \mathrm{sin}^2\theta}{\Sigma |\Lambda |^2}\left(|\Lambda_0 |^2 d\hat{\phi}-\omega ' d\hat{t}\right)^2 , \ \label{melvinkerrmetric}
\end{eqnarray}
where the event horizon does not differ from the Kerr one and if we take $ B=0 $, it will return to Kerr metric.

\section{NHEK black hole and the corresponding Ernst potentials}

Here, we will study the extremal Kerr geometry by using new coordinate \cite{guica,compere,bardeen} on (\ref{kerrmetric}), defined by
\begin{eqnarray}
&&\hat{t}=\frac{2M t}{\lambda}, ~\hat{\phi}=\phi+\frac{t}{\lambda}, ~ \hat{r}=\frac{\lambda M}{y} + M \label{transformkerr} ,\
\end{eqnarray}
and take $ \lambda \rightarrow 0 $ as the extreme limit. Finally by using (\ref{transformkerr}), we obtain
\begin{equation}
ds^2=2J \Omega^2 \left[\frac{-dt^2+dy^2}{y^2}+d\theta^2+\Lambda^2\left(\frac{dt}{y}+d\phi\right)^2\right] , \label{nhekpoincare}\
\end{equation}
which is in the Poincar{\'e}-type coordinates and where
\begin{eqnarray}
 J = M^2, ~ \Omega^2 = \frac{1+\mathrm{cos}^2\theta}{2}, ~\mathrm{and}~ \Lambda = \frac{2\mathrm{sin}\theta}{1+\mathrm{cos}^2\theta}. \nonumber\
\end{eqnarray}
This NHEK geometry is not asymptotically flat. We also find the new Ernst potential along with the using this transformation (\ref{transformkerr}) that only depends on the angular coordinate. The corresponding Ernst potential is
\begin{eqnarray}
\varepsilon = -2M^2\mathrm{sin}^2\theta -\frac{2M^2 \mathrm{sin}^4\theta}{1+\mathrm{cos}^2\theta} + 2iM^2 \mathrm{cos}\theta\left((3-\mathrm{cos}^2\theta)+\frac{\mathrm{sin}^4\theta}{1+\mathrm{cos}^2\theta}\right). \label{ernstpotnhek}\
\end{eqnarray}
where $ \Phi $ is still zero. By using this transformation 
\begin{eqnarray}
d\hat{t} \rightarrow y^2 d\hat{t}, \
\end{eqnarray}
to Poincare-type extremal Kerr geometry, we will have
\begin{equation}
ds^2=2J \Omega^2 \left[-y^2 dt^2+ \frac{dy^2}{y^2}+d\theta^2+\Lambda^2\left(d\phi+yd\hat{t}\right)^2\right] , \label{nheklocal}\
\end{equation}
which is NHEK in the form of general near horizon geometry that is still a vacuum solution of Einstein field equation. Then global NHEK geometry also can be obtained from coordinates transformation, which is defined by
\begin{eqnarray}
\fl y=\frac{1}{r+\sqrt{1+r^2}\mathrm{cos}\tau}, ~ t=y ~ \mathrm{sin}\tau \sqrt{1+r^2},~ \phi =\varphi + \mathrm{ln}\left(\frac{\mathrm{cos}\tau +r~\mathrm{sin}\tau}{1+\mathrm{sin}\tau \sqrt{1+r^2}}\right). \label{nhekglobaltransform}
\end{eqnarray}
By inserting (\ref{nhekglobaltransform}) to (\ref{nhekpoincare}), we find
\begin{equation}
ds^2 = 2J \Omega^2 \left[-(1+r^2)d\tau^2+\frac{dr^2}{1+r^2}+d\theta^2+\Lambda^2\left(d\varphi +r d\tau\right)^2\right] .\label{nhekglobal}
\end{equation}

Now, we can calculate the central charge of NHEK geometry in the stretched horizon formalism. In this formalism, the metric used here is four-dimensional stationary black hole in ADM form \cite{chen}
\begin{eqnarray}
ds^2=-N^2d\hat{t}^2+h_{\hat{r}\hat{r}} d\hat{r}^2+h_{\hat{\theta}\hat{\theta}} d\hat{\theta}^2+h_{\hat{\phi}\hat{\phi}}(d\hat{\phi}+N^{\hat{\phi}}d\hat{t})^2 \label{admkerrmetric}\ 
\end{eqnarray}
The central charge which is related to the (\ref{admkerrmetric}) has general form as
\begin{eqnarray}
c=\frac{3fA_{BH}}{2\pi}, \label{centralcharge} \
\end{eqnarray}
where  $ A_{BH} $ is the horizon area of the black hole and $ \hat{f} $ is a function which is defined by 
\begin{equation}
\hat{f}=\frac{f_2 f_3}{f_1}\bigg|_{\hat{r}=\hat{r}_+}\
\end{equation}
 where
\begin{eqnarray}
& & f_1= \frac{N}{\hat{r}-\hat{r}_+}, ~ f_2=(\hat{r}-\hat{r}_+)\sqrt{h_{\hat{r}\hat{r}}}, ~  f_3=\frac{N^{\hat{\phi}}+\Omega_H}{\hat{r}-\hat{r}_+}.\
\end{eqnarray}
In near horizon, we have
\begin{eqnarray}
f_3|_{r=r_+} = \partial_r N^{\phi}|_{r=r_+} .\
\end{eqnarray}
The event horizon of the NHEK geometry (\ref{nheklocal}) is on $ r_+ =0 $ \cite{chen,chen1,chen2}. So, from the metric (\ref{nheklocal}), we see that
\begin{eqnarray}
f_1 =f_2 = \sqrt{2M^2 \Omega^2}, ~~f_3 =1, \
\end{eqnarray}
and it causes $ \hat{f}=1 $. In addition, the horizon area $ A_{BH} $ is $ 8\pi M^2 $, finally we find the central charge such as
\begin{eqnarray}
c=12M^2 , \label{centralchargekerr}
\end{eqnarray}
where it is equal to the Kerr metric with horizon in $ r_+=M $. For GRS 1915+105, it yields $ c = (2\pm1)\times 10^{79} $ \cite{guica}.

\section{NHEMK Geometry and the corresponding Ernst potentials}
The Harrison transformation does not change the event horizon of the Kerr black hole, so the extremal case happens when $ a=M $.  So, to obtain NHEK geometry, we use these coordinates transformation \cite{siahaan,mei}
\begin{eqnarray}
\hat{t}=\frac{2M^2 t}{\lambda}, ~ \hat{\phi}=\phi+\frac{(1+2B^4M^4)Mt}{(1+B^4M^4)\lambda}, ~  \hat{r}=\lambda y+M. \label{transform1MK} \
\end{eqnarray}
By taking the limit $ \lambda \rightarrow 0 $, we obtain NHEK geometry from (\ref{melvinkerrmetric}) that is defined by
\begin{eqnarray}
\fl ds^2=\Sigma |\Lambda|^2 \left(-y^2 dt^2+\frac{dy^2}{y^2}+d\theta^2 \right)+\frac{A \mathrm{sin}^2\theta |\Lambda_0|^4}{\Sigma |\Lambda |^2}\left(d\phi +\frac{(1-B^4 M^4)}{|\Lambda_0|^2}y dt \right)^2, \ \label{nhemkmetrik1}
\end{eqnarray}
with the corresponding vector potentials, those are $ A_{\mu}dx^{\mu}=(1-B^4M^4)yA(\theta)dt+A(\theta )d\phi$ where
\begin{eqnarray}
A( \theta) = \frac{-4BM \mathrm{cos}\theta}{(1+B^2M^2)^2+(B^2M^2-1)^2 \mathrm{cos}^2\theta} ,\
\end{eqnarray}
where it is little bit different such shown in \cite{siahaan}, however it is still the solution of the Einstein-Maxwell system of equations.
Similar with the NHEK geometry, the Ernst potentials can be obtained from the corresponding coordinates transformation (\ref{transform1MK}). The new Ernst potentials for NHEMK are 
\begin{eqnarray}
\varepsilon ' &=& \frac{-4M^2[1+3M^2B^2+(B^2 M^2-1)^2(2\mathrm{cos}^2\theta-1)]+16iM^2\mathrm{cos}\theta}{3+2B^2 M^2+3B^4 M^4+(B^2 M^2-1)^2(2\mathrm{cos}^2\theta-1)}, \\
\Phi '&=& \frac{2M^2B[1+3M^2B^2+(B^2 M^2-1)^2(2\mathrm{cos}^2\theta-1)]-8iM^2B\mathrm{cos}\theta}{3+2B^2 M^2+3B^4 M^4+(B^2 M^2-1)^2(2\mathrm{cos}^2\theta-1)}, \ \label{ernstpotnhemk}
\end{eqnarray}
which reduce to the Ernst potentials of the NHEK metric by setting the magnetic field parameter $ B $ to $ 0 $. The remained Cartan components, the orthonormal components of the electromagnetic field for the locally nonrotating observer, are only 
\begin{eqnarray}
H_r = \frac{4B[1+6B^2M^2+B^4M^4-(B^2M^2-1)^2 (2\mathrm{cos}^2\theta-1)]}{[3+2B^2M^2+3B^4M^4+(B^2M^2-1)^2 (2\mathrm{cos}^2\theta-1)]^2}, \\
E_r = \frac{16B(1-B^4M^4)\mathrm{cos}\theta}{[3+2B^2M^2+3B^4M^4+(B^2M^2-1)^2 (2\mathrm{cos}^2\theta-1)]^2}, \
\end{eqnarray}
and the angular components vanish.
To obtain NHEMK in global coordinates $(\tau, r, \theta, \varphi)$, we need to use another coordinates transformation, those are
\begin{eqnarray}
y=r+\sqrt{1+r^2}\mathrm{cos}\tau, ~ t=\frac{\mathrm{sin}\tau \sqrt{1+r^2}}{y}, \nonumber\\
\phi =\varphi + \frac{(1-B^4 M^4)}{|\Lambda_0|^2} \mathrm{ln}\left(\frac{1+\mathrm{sin}\tau \sqrt{1+r^2}}{\mathrm{cos}\tau +r~\mathrm{sin}\tau}\right). \
\end{eqnarray}
Finally, the global NHEMK geometry is in the form
\begin{eqnarray}
\fl ds^2 &=&\Sigma |\Lambda|^2 \left[-(1+r^2)d\tau^2+\frac{dr^2}{1+r^2}+d\theta^2 \right]+\frac{A \mathrm{sin}^2\theta |\Lambda_0|^4}{\Sigma |\Lambda |^2}\left[d\varphi +\frac{(1-B^4 M^4)}{|\Lambda_0|^2}r d\tau \right]^2 . \label{nhemkglobal} \
\end{eqnarray}

Stretched horizon formalism can be applied to the general stationary metric in N-dimension. Hence the central charge of the NHEMK geometry can be obtained from using stretched horizon formalism too. But, here we first show the magnetized ADM metric of the rotating black hole by using Harrison transformation. So, the  magnetized ADM form of rotating black hole becomes
\begin{eqnarray}
ds^2=|\Lambda|^{2}(-N^2d\hat{t}^2+h_{\hat{r}\hat{r}} d\hat{r}^2+h_{\hat{\theta}\hat{\theta}} d\hat{\theta}^2)+|\Lambda|^{-2}h_{\hat{\phi}\hat{\phi}}(d\hat{\phi}+N^{\hat{\phi}}d\hat{t})^2 \label{admkerrharrisonmetric}\ .
\end{eqnarray}
Then the function $ \hat{f}' $ can be defined as
\begin{equation}
\hat{f}'=\frac{f^{'}_2 f^{'}_3}{f^{'}_1}\bigg|_{\hat{r}=\hat{r}_+}\
\end{equation}
where
\begin{eqnarray}
&& f^{'}_1= \frac{|\Lambda|N}{\hat{r}-\hat{r}_+}, ~ f^{'}_2=|\Lambda| (\hat{r}-\hat{r}_+)\sqrt{h_{\hat{r}\hat{r}}}, ~ f^{'}_3=\frac{N^{\hat{\phi}'}+\Omega _H ^{'}}{\hat{r}-\hat{r}_+}, \\
&& \nabla N^{\hat{\phi}'}  = |\Lambda|^2 \nabla N^{\hat{\phi}} + \rho f^{-1}\left(\Lambda ^* \nabla \Lambda -\Lambda \nabla \Lambda ^* \right).
 \label{fcentralchargelamabaru}\
\end{eqnarray}
We can also find the function $ \hat{f}' $ in the function of Papapetrou (\ref{papapetrouparam}) and Ernst potentials. So, we have $ f_1, f_2$, and $ f_3 $
\begin{eqnarray}
\fl f^{'}_1= \frac{|\Lambda| \rho}{(\hat{r}-\hat{r}_+)\sqrt{-\mathrm{Re}~\varepsilon}}, ~ f^{'}_2=|\Lambda |(\hat{r}-\hat{r}_+)\sqrt{(\mathrm{Re}~\varepsilon ~ \Delta P^{2})^{-1}}, ~ f^{'}_3=\frac{-\omega^{'}+\Omega_H^{'}}{\hat{r}-\hat{r}_+}. \
\end{eqnarray}

For example, from extremal magnetized Kerr, we can find
\begin{eqnarray}
f'_1 = \sqrt{\frac{\Sigma \Delta}{A}}\frac{|\Lambda|}{\hat{r}-\hat{r}_+}, ~ f'_2 =|\Lambda | \Sigma ^{1/2}, ~ f'_3 = \frac{\Omega_H |\Lambda _0|^2 -\omega ^{'}}{(\hat{r}-\hat{r}_+)|\Lambda _0|^2} . \label{fmelvinkerr} \
\end{eqnarray}
By taking $ \hat{r}=\hat{r}_+ $ on (\ref{fmelvinkerr}), we obtain
\begin{eqnarray}
f'=\frac{1-B^4 M^4}{1+B^4 M^4}. \
\end{eqnarray}
The horizon area of the extremal magnetized Kerr black hole is defined by
\begin{eqnarray}
A_{BH}=\int \sqrt{|\hat{\gamma} |} d\theta d\phi, \label{areabh}
\end{eqnarray}
with
\begin{eqnarray}
|\hat{\gamma} | = 4 M^4  \mathrm{sin}^2 \theta (1+B^4 M^4)^2.\
\end{eqnarray}
Then by integrating (\ref{areabh}), the area becomes
\begin{equation}
A_{BH}=8\pi M^2 (1+B^4 M^4). \
\end{equation}
After finding $ f' $ and $ A_{BH} $, we can find the central charge (\ref{centralcharge}) to  be
\begin{equation}
c = 12M^2 (1-B^4 M^4), \
\end{equation}
where it reduces to the central charge associated to NHEK geometry if there is no external magnetic field. But, strong magnetic field, $|B| > 1/M $, can also make the central charge to be negative and this is not unitary \cite{francesco}. So we need $ B $ to be weak in order to obtain the positive central charge. By using another formula given in \cite{hartman}, we can find the same central charge (\ref{centralcharge}) where it is defined as
\begin{eqnarray}
c_{grav} = 3k \int^{\pi}_0 \left[\Gamma(\theta)\alpha(\theta)\gamma(\theta)\right]^{1/2} d\theta , \label{stromingcharge} \
\end{eqnarray}
which is related to the general form of near horizon extremal metric
\begin{eqnarray}
ds^2 = \Gamma(\theta)\left[ -r^2 dt^2 +\frac{dr^2}{r^2} +\alpha(\theta)d\theta ^2 \right]+\gamma(\theta)(d\phi+krdt)^2 ,\
\end{eqnarray}
which is equivalent with the near horizon extremal metrics (\ref{nheklocal}) and (\ref{nhemkmetrik1}).

Naively, we can say that if we can find the central charge associated to the extremal Kerr from the extremal magnetized Kerr, indeed, we can reverse the process. We mean that the central charge associated to the extremal magnetized Kerr can be also obtained from the extremal Kerr by Harrison transformation such in (\ref{fmelvinkerr}). We find that the key is on the function $ f'_3 $ because the Harrison transformation parameter $ \Lambda $ in $ f'_1 $ and $ f'_2 $ will cancel each others. We actually can simplify $ f'_3 $ in (\ref{fmelvinkerr}) for the extremal case that satisfies
\begin{eqnarray}
f'_3 = \frac{d\omega '}{d\omega |\Lambda_0| ^2} f_3 = -|\Lambda_0| ^{-2} \frac{ d\omega '}{dr} f_3  , \
\end{eqnarray}
by comparing the Papapetrou metric (\ref{papapetrou}) with (\ref{nhekglobal}) and (\ref{nhemkglobal}) where in this case $ \omega '=  -(1-B^4M^4)r $ and $\omega = -r $. The area of the NHEMK is also related to the NHEK that satisfies
\begin{eqnarray}
A'_{BH} = A_{BH} |\Lambda_0| ^2 , \
\end{eqnarray}
where $ A'_{BH} $ is the area of the NHEMK black hole.
Again, $ |\Lambda_0|^2 $ emerges because of the reason to remove the conical singularity. So, this factor will not change the transformation of the central charge. Hence the central charge satisfies this relation
\begin{eqnarray}
c'=c \frac{d\omega '}{d\omega} .\
\end{eqnarray}
Note that this is not only can be applied to the NHEMK geometry but also possibly to the extremal case of the more general black hole, such as Near Horizon Extremal Magnetized Kerr-Newman that are prepared for the next paper.

\section{Summary}

Black hole solutions are related to the Ernst potentials which are the solutions of the Ernst equations which are equivalent to the Einstein field equation. In Ernst formalism, we use general rotating black hole metric called by Papapetrou metric. This metric can be applied to the case of Kerr metric. On the other hand, black hole solutions, especially on the near horizon extremal case, are conjectured to be  holographic dual to the conformal field theory, or so-called by Kerr/CFT correspondence. This discovery has been quite successful as the tools to find the microscopic origin of the Bekenstein-Hawking entropy. 
The entropy of the extremal black holes is related to the central charge that appears from the Cardy formula. To obtain the corresponding central charge, we can adopt the method by Brown and Henneaux which is associated to $ \textrm{AdS}_3 $ spacetime.

The tools of Kerr/CFT correspondence are also applied to the magnetized Kerr black holes and also have a warped and twisted product of $ \textrm{AdS}_2 \times \textrm{S}^2 $ such as the extremal Kerr one. Mathematically, this magnetized geometry is obtained from the Harrison transformation. By applying Harrison transformation to the metric, the central charge is also found to depend on the magnetic fields parameter and also the Ernst potentials that depend only on the angular coordinate. These Ernst solutions are still the solution of the Ernst equations. Then the vector potentials of NHEMK geometry are also obtained as the electrovacuum solution of the Einstein-Maxwell system of equations which correspond to the Cartan components. The central charge of NHEMK geometry can be obtained by using stretched horizon formalism, similar to the NHEK geometry. We show the stretched horizon formalism in the function of the Ernst-Papapetrou potentials and also for the magnetized ADM metric one. Note that by taking $ B $ to be zero will reduce to the case of NHEK. Then we show the transformation of the central charge associated to NHEK geometry to NHEMK geometry in the function of Ernst and Papapetrou potentials.

The transformation of the central charge associated to NHEK geometry to NHEMK geometry in the function of Ernst and Papapetrou potentials possibly can be generalized. It means that this kind of transformation, is also satisfied by the more general black hole, i.e. Kerr-Newman, especially in the near horizon extremal case. If this is proved, we have find a simpler way to find the central charge of any black hole from the simpler black hole one as we have shown in this paper.

\section*{Acknowledgments}

We gratefully acknowledge support by Riset PMDSU 2016 from Ministry of Research, Technology, and Higher Education of the Republic of Indonesia. M F A R S would like to thank Haryanto M Siahaan for giving us the topic and the stimulating discussion. M.F.A.R.S also thanks all members of Theoretical Physics Laboratory, Institut Teknologi Bandung for the valuable support.


%
%
%
%

\section*{References}
\nocite{*}
\bibliographystyle{aipnum-cp}
\bibliography{sample}

\begin{thebibliography}{1}

\bibitem{ernst}
Ernst F J 1967 \textit{Phys. Rev.} \textbf{167} 1175

\bibitem{ernst1}
Ernst F J 1967 \textit{Phys. Rev.} \textbf{168} 1415

\bibitem{reina}
Reina C and Treves A 1976 \textit{Gen. Relativ. Grav.} \textbf{7} 817

\bibitem{brown}
Brown J D and Henneaux M 1986 \textit{Commun. Math. Phys.} \textbf{104} 207


\bibitem{elvis}
Elvis M, Risaliti G and Zamorani G 2002 \textit{Astrophys. J.} \textbf{565}, L75

\bibitem{harrison}
Harrison B K 1976 \textit{J. Math. Phys.} \textbf{9}, 1744

\bibitem{siahaan}
Siahaan H M 2016 \textit{Class. Quantum Grav.} \textbf{33} 155013

\bibitem{astorino}
Astorino M 2015 \textit{Phys. Lett.} B \textbf{751} 96

\bibitem{astorino2}
Astorino M 2015 \textit{J. High Energy Phys.} JHEP10(2015)016

\bibitem{garousi}
Garousi M R and Ghodsi A 2010 \textit{Phys. Lett.} B \textbf{687} 79

\bibitem{chen}
Chen B and Zhang J J 2012 \textit{Nucl. Phys.} B \textbf{856} 449

\bibitem{hartman}
Hartman T, Murata K, Nishioka T and Strominger A 2009 \textit{J. High Energy Phys.} JHEP04(2009)019

\bibitem{kerr}
Kerr R P 1963 \textit{Phys. Rev. Lett.} \textbf{11} 237

\bibitem{aliev}
Aliev A N and Galtsov D V 1988 \textit{Astrophys. Space Sci.} \textbf{143} 301

\bibitem{hiscock}
Hiscock W A 1981 \textit{J. Math. Phys.} \textbf{22} 1828

\bibitem{guica}
Guica M, Hartman T, Song W and Strominger A 2008 \textit{Phys. Rev} D \textbf{80} 124008

\bibitem{compere}
Compere G 2012 \textit{Living Rev. Rel.} \textbf{15} 11

\bibitem{bardeen}
Bardeen J and Horowitz G T 1999 \textit{Phys. Rev.} D \textbf{60} 104030

\bibitem{chen1}
Chen B and Chu C S 2010 \textit{J. High Energy Phys.} JHEP05(2010)004

\bibitem{chen2}
Chen B, Long J and Zhang J J 2010 \textit{Phys. Rev.} D \textbf{82} 104017

\bibitem{mei}
Mei J 2010 \textit{J. High Energy Phys.} JHEP04(2010)005

\bibitem{francesco}
Di F P, Mathieu P and S{\'e}n{\'e}chal D 1997  \textit{Conformal Field Theory} (New York: Springer-Verlag)

\end{thebibliography}

\end{document}